\documentclass[a4paper,fleqn,usenatbib]{mnras}
\usepackage{times,txfonts}
\usepackage[T1]{fontenc}
\usepackage{ae,aecompl}
\usepackage{graphicx}

\newcommand{\MSUN}{\mbox{$\mathrm{M_{\odot}}$}}

\title[FGK stars with white dwarf companions]{The White Dwarf Binary Pathways
  Survey I: A sample of FGK stars with white dwarf companions}

\author[S. G. Parsons et al.]{S.~G.~Parsons$^{1}$\thanks{steven.parsons@uv.cl},
A.~Rebassa-Mansergas$^{2}$,
M.~R.~Schreiber$^{1,3}$,
B.~T.~G{\"a}nsicke$^{4}$,
\newauthor
M.~Zorotovic$^{1}$
and J.~J.~Ren$^{5}$
\\
$^{1}$ Instituto de F{\'i}sica y Astronom{\'i}a, Universidad de
Valpara{\'i}so, Avenida Gran Bretana 1111, Valpara{\'i}so, 2360102, Chile\\
$^{2}$ Departament de F\'isica, Universitat Polit\'ecnica de Catalunya,
c/Esteve Terrades 5, 08860 Castelldefels, Spain\\
$^{3}$ Millenium Nucleus "Protoplanetary Disks in ALMA Early Science",
Universidad de Valparaiso, Valparaiso 2360102, Chile\\
$^{4}$ Department of Physics, University of Warwick, Coventry CV4 7AL, UK\\
$^{9}$Department of Astronomy, Peking University, Beijing 100871,
P.\,R.\,China}

\date{Accepted 2016 August 23. Received 2016 August 16; in original form 2016 March 21}

\pubyear{2016}

\begin{document}
\label{firstpage}
\pagerange{\pageref{firstpage}--\pageref{lastpage}}
\maketitle

\begin{abstract}
The number of spatially unresolved white dwarf plus main-sequence star
binaries has increased rapidly in the last decade, jumping from only $\sim$30
in 2003 to over 3000. However, in the majority of known systems the companion
to the white dwarf is a low mass M dwarf, since these are relatively easy to
identify from optical colours and spectra. White dwarfs with more massive FGK
type companions have remained elusive due to the large difference in optical
brightness between the two stars. In this paper we identify 934 main-sequence
FGK stars from the Radial Velocity Experiment (RAVE) survey in the southern
hemisphere and the Large Sky Area Multi-Object Fiber Spectroscopic Telescope
(LAMOST) survey in the northern hemisphere, that show excess flux at
ultraviolet wavelengths which we interpret as the likely presence of a
white dwarf companion. We obtained Hubble Space Telescope ultraviolet
spectra for nine systems which confirmed that the excess is indeed caused, in
all cases, by a hot compact companion, eight being white dwarfs and one a hot
subdwarf or pre-helium white dwarf, demonstrating that this sample is very
clean. We also address the potential of this sample to test binary evolution
models and type Ia supernovae formation channels. 
\end{abstract}

\begin{keywords}
binaries: close -- stars: white dwarfs -- stars: early-type -- stars: evolution
\end{keywords}

\section{Introduction}

Around 25 per cent of all binary stars are born close enough that they will
eventually interact with each other when the more massive member of the binary
evolves off the main-sequence and becomes a giant star
\citep{willems04}. Very often this results in material flowing from the
expanding giant onto the lower mass star, which cannot incorporate 
it fast enough and therefore the material overflows the lower-mass star's
Roche lobe, leading to the entire binary being engulfed in what is known as a
common envelope (CE, \citealt{paczynski76,webbink84}). Inside this
envelope the stars are subject to strong frictional forces leading to a loss
of angular momentum and orbital energy, hence the two stars spiral inward
towards each other, ejecting the envelope as they do so. What emerges from the
CE is a close binary, known as a post common-envelope binary (PCEB),
containing the core of the giant star (which will become a white dwarf) and
its main-sequence star companion. Depending upon the stellar masses and final
separation after the CE stage, the binary may go on to experience a second CE
stage when the main-sequence star evolves, resulting in a double degenerate
binary system \citep{webbink84,nelemans01,toonen12}. Alternatively, the stars
may have emerged from the CE close enough to each other that angular momentum
loss will drive them in to contact before the main-sequence star has evolved,
often leading to the formation of a cataclysmic variable (CV) system.

PCEBs are important objects since they are likely the progenitors of supernova
Ia (SN\,Ia). While it is generally established that SN\,Ia are the result of
the thermonuclear ignition of a carbon-oxygen core white dwarf, there is not
yet a consensus on the pathway leading to the explosion. In the classic double
degenerate scenario the merger of two white dwarfs with a combined mass above
the Chandrasekhar limit leads to a SN\,Ia explosion
\citep{tutukov79,webbink84}. This channel has the advantage that it can
produce systems with both long and short time delays and explain the lack of
hydrogen and helium lines in most SN\,Ia spectra \citep{branch95}. Moreover,
there is both theoretical and observational evidence for a large population of
close double white dwarf binaries
\citep{yungelson94,nelemans05,napiwotzki01}. However, there is some 
uncertainty as to whether the merger leads to an explosion or 
rather a core collapse and the formation of a neutron star instead
\citep{nomoto91} and at present the situation remains uncertain
\citep{yoon07}. Moreover, a range of explosion masses are possible with
  the merger of two white dwarfs of arbitrarily mass, which would cause a
  spread in the luminosities of SN\,Ia, affecting their use as standard
  candles. In the alternative scenario, known as the single degenerate
channel, a white dwarf stably accretes material from a non-degenerate
companion star until it grows to the Chandrasekhar mass and explodes
\citep{iben84}. In this case the progenitor systems are thought to be white
dwarfs accreting from main-sequence or giant stars at a sufficiently high rate
that hydrogen is stably burnt on the surface of the white dwarf, avoiding nova
explosions \citep{shara77,iben82}. Such systems are observed in the Galaxy and
Magellanic Clouds \citep{vdheuvel92,kahabka97} and are known as supersoft X-ray
sources (SSSs). However, in order to sufficiently grow the white dwarf, models
often require fine-tuning many uncertain input parameters such as the
accretion rate \citep{hillebrandt00,wang12}. Furthermore, the single
degenerate channel has difficulty in creating SN\,Ia with time delays larger
than a few Gyr due to the short main-sequence lifetimes of their
  progenitors \citep[e.g.][]{claeys14}. 

In recent years the idea of sub-Chandrasekhar or
double-detonation models have become popular. In this model the white dwarf
has accumulated a substantial surface layer of helium. When this layer ignites
it can send a strong shock wave into the core of the white dwarf, compressing
it and triggering a secondary detonation of the core, leading to a SN\,Ia
explosion \citep{fink07,sim10,kromer10}. This can be achieved via the double
degenerate route, with the merger of a carbon-oxygen core and a helium core
white dwarf, or the single degenerate route, with the white dwarf accreting
material from a helium-rich companion star. This idea has the advantage that
there are a sufficient number of such binaries to explain the birthrate of
SN\,Ia \citep{ruiter09}, although it is difficult to explain the observed
similarities in most SN\,Ia via this model \citep[e.g.][]{branch95}. In
virtually every model for the formation of SN\,Ia the progenitor systems are
binaries that have passed through at least one CE event (two in the case of
the double degenerate channel), hence PCEBs are a valuable population for
testing the ideas of SN\,Ia formation. In particular, PCEBs containing white
dwarfs with massive FGK-type main-sequence star companions are both potential
SN\,Ia progenitors (via the single degenerate channel) and an intermediate
stage (via the double degenerate channel) and therefore are a particularly
useful population. A large sample of such systems can even be used to estimate
the relative number of systems evolving through the double degenerate and the
single degenerate channels.

The number of PCEBs with late-type companions has rapidly increased in the
last few years thanks mainly to the Sloan Digital Sky Survey (SDSS) and its
spectroscopic database
\citep{raymond03,silvestri06,rebassa07,rebassa12,li14}. With 
optical spectra alone it is possible to detect even very cool white dwarfs
next to M-type stars \citep[e.g.][]{parsons12} and hotter white dwarfs are
still detectable with companions up to very-late K ($\sim$5 per cent of SDSS
white dwarf plus main-sequence star binaries contain K star components,
\citealt{rebassa16}). However, detecting PCEBs with earlier-type companions is
more chalenging because these hotter, larger main-sequence stars completely
outshine white dwarfs in the optical, implying that a combination of
ultraviolet (UV) and optical data is required. 

Several attempts have been made to identify white dwarfs with FGK-type
main-sequence star companions \citep[e.g.][]{maxted09}, which have succeeded
in finding these systems, but have failed to produce large numbers of
them. For example, SDSS FGK stars are generally located at large enough
  distances that all but the very hottest white dwarfs will go undetected in
  UV surveys, severely reducing the sample size. Previous studies of the
far-UV and soft X-ray spectra of main-sequence stars have revealed the
presence of white dwarf companions
\citep{landsman93,barstow94,vennes95,christian96,landsman96,vennes97,burleigh97,burleigh98,burleigh98b,burleigh99}, 
but the vast majority of these systems are visually resolved binaries that
were formed far enough apart that they avoided a CE event. To
date, only 7 PCEBs with early-type main-sequence star companions have been
identified and had their orbital periods measured
\citep{nelson70,simon85,berghofer00,vennes98,kruse14,parsons15}. The
difficulty in detecting white dwarf companions to FGK stars is not only due to
the extreme optical contrast between the two stars (typically the white dwarf
will only contribute $\sim$1 per cent of the potical light), but also due to
the fact that many of the largest samples of FGK stars (e.g. from the SDSS)
are lacking in bright (V<12) members because they saturate in modern survey
data. This means that these samples contain fainter stars that are located at
large distances. For example, the saturation limits for SDSS (about 14 mag in
$g$, $r$ and $i$ and about 12 mag in $u$ and $z$, \citealt{chonis08}) mean
that main-sequence FGK stars are generally located at more than 0.5\,kpc from
Earth \citep{bilir09}. At these large distances only the hottest (hence
youngest and most short-lived) white dwarf companions will be bright enough to
be detected in the UV, severely limiting the sample size (only $\sim$2 per
cent of UV-excess FGK stars in SDSS are the result of a white dwarf companion,
\citealt{smith14}). Therefore, searching for PCEBs with early-type companions
requires a sample of bright, nearby FGK stars with UV data.

To this end, we have started a project that uses data from surveys that
target bright FGK stars in combination with UV data from the Galaxy Evolution
Explorer (GALEX) database in order to detect main-sequence stars with
substantial UV-excess colours that we interpret as the result of a nearby
white dwarf companion. Follow-up ground-based spectroscopic observations can
then be used to distinguish between the close PCEBs and wide systems that
avoided a CE event. The final aim of this project is to establish the first
large sample of PCEBs with early-type components, suitable for population
synthesis studies and useful for testing models of SN\,Ia formation. The first
result from this project \citet{parsons15} was the identification of
TYC\,6760-497-1 as a white dwarf plus F-star in a 12 hour binary, which is the
first known progenitor of a SSS, i.e. the white dwarf will grow in mass in its
future evolution. In this paper we present the initial step in our project, the
selection process and sample of UV-excess FGK stars. The spectroscopic
follow-up data and identification of PCEB/wide systems will be presented in a
future paper. We have also obtained Hubble Space Telescope (\textit{HST}) UV
spectroscopy of 9 of these stars that unambiguously prove that the excesses
are indeed due to white dwarfs and we present these data in this paper.

\section{Sample selection}

In this section we outline the methods we used to select our sample of FGK
main-sequence stars with UV-excesses. Note that these stars may be close,
short-period PCEBs (the main targets of this project), or
they may be wide\footnote{Throughout this paper, the term \textit{wide
    binaries} refers to systems that did not undergo CE evolution. Such
  systems may have orbital separations of as little as $\sim$10 AU, and can
  therefore remain unresolved in the available imaging, hence we note that
  \textit{wide} should not be mistaken for \textit{spatially resolved}.} white
dwarf plus FGK binaries, or they may be a chance alignment of an FGK star with
a foreground or background white dwarf. Other potential sources of
contamination are discussed in Section~\ref{sec:contam}. 

We used data from two surveys, the Radial Velocity Experiment (RAVE) survey in
the southern hemisphere and the Large Sky Area Multi-Object Fiber
Spectroscopic Telescope (LAMOST) survey in the northern hemisphere. We begin
by discussing these two samples separately. 

\subsection{The southern hemisphere sample}

In order to identify main-sequence FGK stars with UV-excesses in the southern
hemisphere sky, we used data from the RAVE
survey\footnote{https://www.rave-survey.org} data release 4
\citep{kordopatis13}. RAVE is a magnitude limited survey of 
randomly selected southern hemisphere stars spanning 9$<I<$12. It has
collected spectra for over 425,000 stars covering the infrared calcium triplet
(8410$-$8795\AA) with a resolution of $R\sim7500$ using the 6dF facility on
the 1.2m Schmidt Telescope at Anglo-Australian Observatory in Siding Spring,
Australia. The data are then used to determine atmospheric parameters
(effective temperature, surface gravity, metallicity) as well as a measurement
of the radial velocity of the star. For a more detailed description of the
RAVE survey we refer the reader to \citet{kordopatis13} and references
therein. 

\begin{figure}
\begin{center}
 \includegraphics[width=\columnwidth,bb=10 10 500 400,clip]{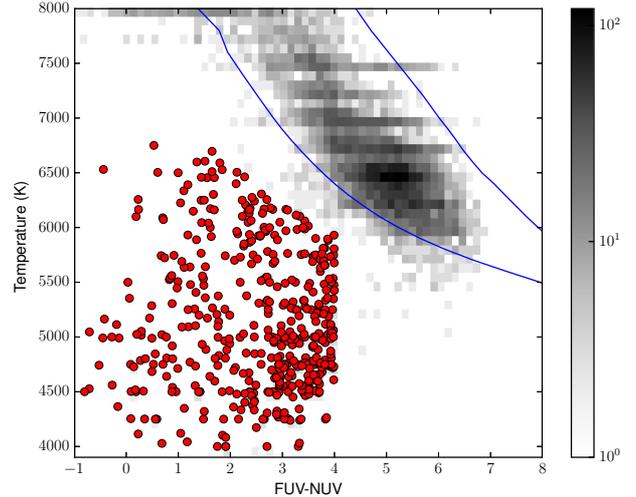}
 \caption{Distribution of the 23,484 RAVE dwarf stars ($\log{g}>3.5$) with
   both a GALEX FUV and NUV detection. The temperatures are taken directly
   from the RAVE DR4 catalogue. Shown in blue are two theoretical tracks
   computed from PHOENIX atmospheric models, one with a high metallicity and
   high surface gravity ($\log{Z}=+1$ $\log{g}=5.0$, top curve) and one with a
   low metallicity and low surface gravity ($\log{Z}=-3$ $\log{g}=3.5$, bottom
   curve). These represent extreme limits on the colours of main-sequence
   stars. The vast majority of RAVE stars fall within these extreme models, as
   expected. The red points mark the 430 targets that fall within our cut
   (FUV$-$NUV$<$4, T$>$4000\,K and FUV$-$NUV at least 1.5 magnitudes bluer than
   the bottom model).} 
 \label{fig:ravedis}
 \end{center}
\end{figure}

We selected all the main-sequence (non-giant, $\log{g}>3.5$) stars in the RAVE
survey with effective temperatures between 4,000\,K and 7,000\,K
(corresponding to F, G and K stars) and cross correlated them with UV data
from the GALEX survey \citep{martin05}, selecting all the sources with both
far-UV (FUV) and near-UV(NUV) measurements (with errors below 0.2 mag and no
artifacts associated to their photometry), resulting in 23,484 main-sequence
stars with good UV photometry. The distribution of effective temperatures and
FUV$-$NUV colours of these stars are shown in Figure~\ref{fig:ravedis}. 

We used the extensive library of PHOENIX stellar synthetic spectra from
\citet{husser13} to determine the UV colours of single main-sequence stars
over a range of temperatures by convolving the synthetic spectra with the
GALEX FUV and NUV filter profiles. We used synthetic spectra spanning surface
gravities of 3.5$<\log{g}<$5.0 and metallicities of -3.0$<\log{Z}<$1.0, but
did not include any alpha element enhancement or depletion. We found that the
UV colours of main-sequence stars are strongly affected by both their surface
gravity (due to the surface gravity dependence in the strength of the Balmer
jump, and the associated suppression of NUV flux) and their metallicity (which
causes variations in the UV opacity), leading to a spread in the FUV$-$NUV
colour of almost 3 magnitudes for stars with the same effective
temperature\footnote{activity is likely to further increase the intrinsic
  scatter, particularly for the lower mass stars}. In Figure~\ref{fig:ravedis}
we plot two tracks showing the UV colours of the bluest and reddest
main-sequence stars within our parameter range. We expect single FGK
main-sequence star from RAVE to have UV colours somewhere between these two
tracks. Indeed, Figure~\ref{fig:ravedis} shows that the FUV$-$NUV colours of
the vast majority of FGK stars in RAVE fall within these two extremes
($\simeq$99 per cent). However, there are also a small number of
stars with substantially bluer UV colours which may be caused by the
presence of a nearby white dwarf companion. We flag a star as having a
UV-excess if it has a FUV$-$NUV colour at least 1.5 magnitudes bluer than the
bluest main-sequence star models for its effective temperature(the bottom
curve in Figure~\ref{fig:ravedis}). We also place an upper limit on the
FUV$-$NUV colour of 4, to avoid including any stars with erroneous RAVE
temperatures.

In total our sample contains 430 stars with UV-excesses, these
are highlighted in Figure~\ref{fig:ravedis} as red points and are detailed in
Table~\ref{tab:ravetargs} in the appendix. The completeness and any
  sources of bias in our sample will be addressed in a future paper.

We checked our sample of UV-excess FGK stars against the list of white dwarf's
with early-type companions from \citet{holberg16}. We found that only 4 of
the 98 objects from the \citet{holberg16} catalogue are in the RAVE survey,
and only one of these, WD\,0354-368 (53,000K white dwarf plus G2 star with
a sky projected separation of $\sim$0.5\arcsec), has GALEX FUV and NUV
measurements (most objects in the \citet{holberg16} catalogue are too bright
for GALEX). However, this star did make it into our cut.  

\subsection{The northern hemisphere sample}

\begin{figure}
\begin{center}
 \includegraphics[width=\columnwidth,bb=10 10 500 400,clip]{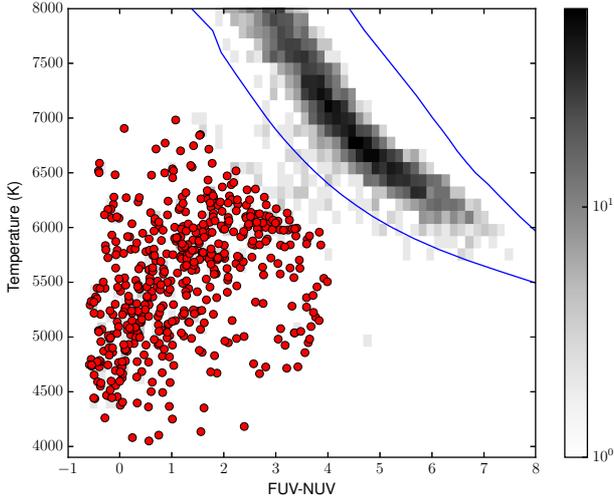}
 \caption{The UV colours and temperatures of LAMOST main-sequence FGK
   stars. We plot the same two PHOENIX models shown in
   Figure~\ref{fig:ravedis}, which represent the most extreme colours for 
   main-sequence stars. Stars flagged as having UV-excesses are shown as red
   points and have UV colours at least 1.5 magnitudes bluer than the bottom
   model. The LAMOST sample does not suffer from the same discretisation
     issue seen in the RAVE sample, hence why it appers smoother.}
 \label{fig:lamostdis}
 \end{center}
\end{figure}

The Large sky Area Multi-Object fiber Spectroscopic Telescope (LAMOST)
is a $\simeq$4 meter effective aperture Meridian reflecting Schmidt
telescope with a 5 degrees diameter field of view
\citep{cuietal12-1, luoetal12-1, yuanetal15-2}. As a dedicated-survey
telescope, LAMOST makes use of spectral plates to observe 4,000
spectroscopic plus calibration targets in one single exposure, equally
distributed among 16 fiber-fed spectrographs. The spectra cover the
entire optical wavelength range ($\simeq$3,700$-$9,000$\AA$) at
a resolving power of $\sim$2,000. The two main surveys carried out
by LAMOST are LEGAS (the LAMOST Extra-Galactic Survey of
galaxies), that aims at studying the large scale structure of the
Universe \citep{zhaoetal12-1}, and LEGUE (the LAMOST Experiment for
Galactic Understanding and Exploration), that aims at obtaining millions
of main-sequence star spectra to study the structure and evolution of
the Milky Way \citep{dengetal12-1}. Unlike the SDSS, the LEGUE survey
follows a target selection algorithm that is not restricted by a lower
magnitude limit \citep{zhaoetal12-1, carlinetal12-1}, thus allowing
collection of a large number of main-sequence star spectra that are bright
enough to search for white dwarf companions. This, together with the
fact that the LAMOST is located in the Xinglong observing station of
China, allows us to complement our search of white dwarf plus FGK 
binaries in the northern hemisphere.

From 2012 September, the LAMOST has been performing a five-year
regular survey. Before that there was a one-year pilot survey followed
by a two-year commissioning survey. The current data product of the
LAMOST that is publicly available is data release 1 (DR1). DR1
contains a total of 2,204,696 spectra, 1,944,392 of which are
catalogued as main-sequence stars by the LAMOST pipeline
\citep{luo15}. Of these, 794,371 are flagged as F, G or K
main-sequence stars and all of them have effective
temperature, surface gravity and metallicity values derived via the
LAMOST stellar parameter pipeline \citep{wuetal14-1, xiangetal15-1}. We
selected only those objects having spectra with a signal-to-noise ratio
of greater than 10, to ensure that the stellar parameters are robust.

\begin{figure}
\begin{center}
 \includegraphics[width=\columnwidth]{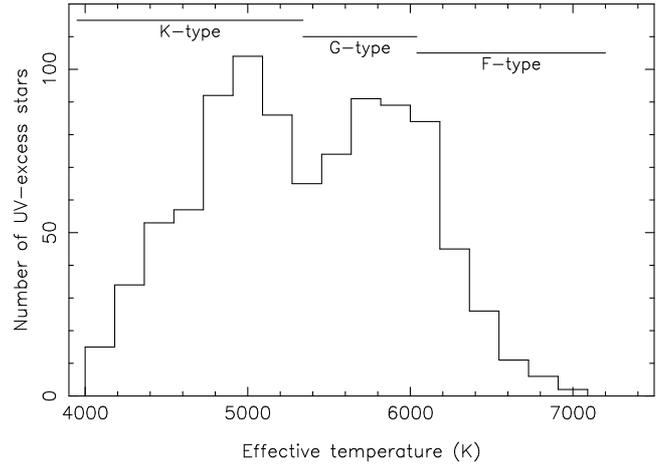}
 \caption{The distribution of temperatures of the UV-excess stars in our
   combined RAVE+LAMOST sample. The approximate temperature ranges for
   standard ($\log{g}=5$) F, G and K type main-sequence stars are indicated at
   the top of the plot.} 
 \label{fig:tdist}
 \end{center}
\end{figure}

We cross-correlated the $\sim$800,000 LAMOST FGK main-sequence stars
with GALEX and found NUV and FUV detections for 4632
objects. In all cases the ultraviolet magnitude errors were below
0.2\,mag and no artifacts were associated to their photometry. As for the RAVE
sample, we selected white dwarf plus FGK binary candidates based on the
detection of an ultraviolet excess, defined as being, for a given temperature,
1.5 magnitudes bluer than the bluest PHOENIX main-sequence colour track (see
Figure~\ref{fig:lamostdis}). The total number of LAMOST FGK stars with
UV-excesses selected in this way is 504. These objects are detailed in
Table~\ref{tab:lamosttargs} in the appendix. As with the RAVE sample, there
was only one previously known system from the \citet{holberg16} catalogue
in the LAMOST database with good UV photometry, WD\,1024+326 (an unresolved
41,000K white dwarf plus G5 system) and this system made it into our cut.

\section{Properties of the sample}

In total we have found 934 (430 from RAVE, 504 from LAMOST) main-sequence FGK
stars that show a UV-excess that may be indicative of the presence of a nearby
white dwarf companion. In Figure~\ref{fig:tdist} we show the distribution of
temperatures (and approximate spectral types) for these stars. Approximately
half our sample consists of K-type stars (471 objects), whilst one third
are G-type stars (307 objects) and the remainder are F-type stars(156
objects).

\begin{figure}
\begin{center}
 \includegraphics[width=\columnwidth]{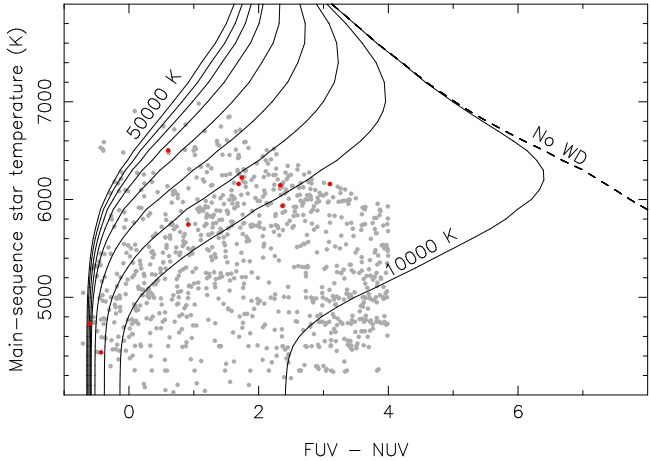}
 \caption{The effect of a white dwarf companion on the UV colours of
   main-sequence stars. The dashed line shows the FUV$-$NUV of $\log{g}=4.5$,
   solar metallicity main-sequence stars as computed from PHOENIX models. The
   solid lines show how the colours of these stars are changed due to the
   presence of a $\log{g}=8.0$ white dwarf companion at the same distance. The
   white dwarf temperatures increase in steps of 5000\,K from right to
   left. The gray points correspond to our RAVE/LAMOST UV-excess objects and
   the red points are those for which we have obtained \textit{HST} UV
   spectra.} 
 \label{fig:wdtemp}
 \end{center}
\end{figure}

\begin{figure}
\begin{center}
 \includegraphics[width=\columnwidth]{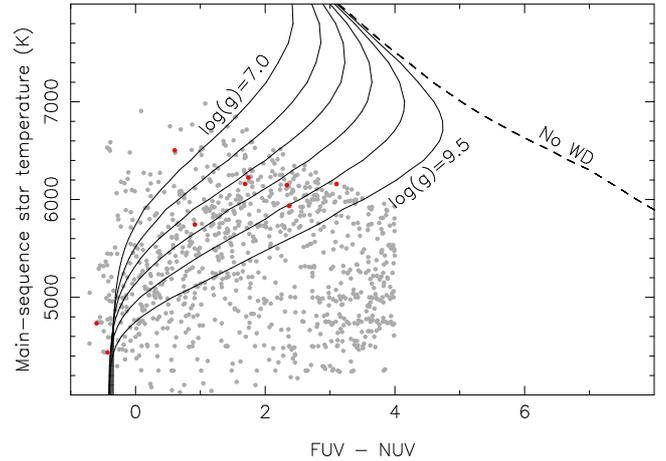}
 \caption{Same as Figure~\ref{fig:wdtemp} but placing 20,000K white dwarfs
   with varying surface gravity next to $\log{g}=4.5$, solar metallicity
   main-sequence stars. The white dwarf surface gravity increases from left
   to right in steps of 0.5 dex.}  
 \label{fig:wdlogg}
 \end{center}
\end{figure}

To estimate the range of white dwarf parameters that our sample can probe, we
combined the PHOENIX model spectra of main-sequence stars with synthetic
spectra of white dwarfs from \citet{koester10}. We used the Torres relation
\citep{torres10} to estimate the radius of the main-sequence star, and the
cooling models of \citet{holberg06} to estimate the radius of the white
dwarf. We then combined the spectra of both stars and computed the resulting
NUV$-$FUV colour. In Figure~\ref{fig:wdtemp} we show the effects on the
NUV$-$FUV colour of varying the temperature of the white dwarf, while keeping
its surface gravity fixed at $\log{g}=8.0$. The figure shows that our sample
could potentially contain systems with white dwarfs as cool as 10,000K next to
K stars, 15,000K next to G stars and 20,000K next to F stars, hence covering a
large range of cooling times. However, the effects of surface gravity are also
important. Figure~\ref{fig:wdlogg} shows the results of keeping the
temperature of the white dwarf fixed at 20,000K and varying it's surface
gravity. The figure shows that surface gravity only has a very minor effect on
the detectability of a white dwarf next to a K star and only a moderate effect
next to a G star. However, when placed next to an F star the surface gravity
of the white dwarf is important for its detectability. High surface gravity
(hence more massive and smaller) white dwarfs only contribute a small fraction
of the overall UV flux against an F star and hence these binaries would not be
detected in our cut. The situation is better at higher temperatures, above
$\sim$25,000K any white dwarf should be detected, regardless of its surface
gravity. Therefore, our sample of systems with F stars likely contains only
the younger members. Fortunately, given the shorter main-sequence lifetimes of
F stars (compared to later spectral types), there is unlikely to be a large
population of these stars with cool ($T_\mathrm{eff}<20,000$K) white dwarf
companions.

\begin{figure}
\begin{center}
 \includegraphics[width=\columnwidth]{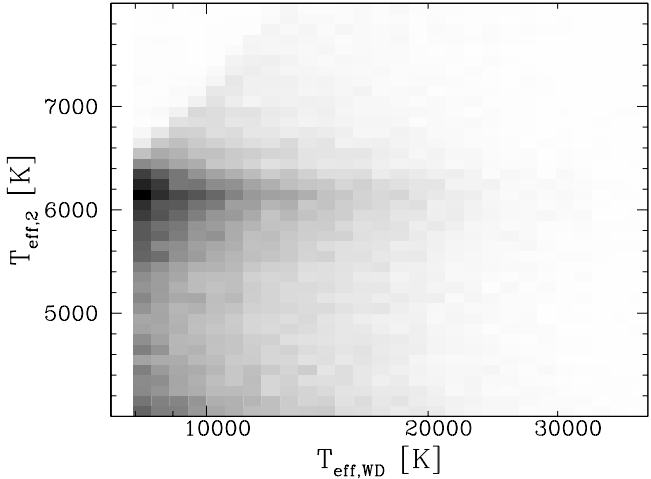}
 \caption{Simulated distribution of white dwarf and main-sequence star
   temperatures. Our sample of UV-excess objects does not cover systems
   containing cool ($T_\mathrm{eff}<20,000$K) white dwarfs and hot (F-type,
   $T_\mathrm{eff}>=6,500$K) main-sequence stars. However, this plot shows
   that the population of systems of this type is very minor and therefore we
   are not missing a large number of systems. The intensity of the grey-scale
   represents the density of simulated objects in each bin, on a linear scale,
   and normalized to one for the bin that contains most systems.}  
 \label{fig:simu}
 \end{center}
\end{figure}

To reinforce this point we simulated the evolution of a population of binaries
with FGK-type secondary stars in order to see the distribution of white dwarf
temperatures in the resulting population. The simulations were performed
in the same way as in \citet{zorotovic14b} assuming a flat initial-mass-ratio
distribution, a common envelope efficiency of 0.25, solar metallicity and
without any contribution from recombination energy. Our results include all
type of detached white dwarf plus FGK systems, not only PCEBs, but also wide
system. The binary star evolution code from \citet{hurley02} uses very old
models \citep{mestel52} to calculate the white dwarf effective
temperature. Therefore, more accurate values were calculated following
\citet{zorotovic16} i.e. by using the cooling tracks from \citet{althaus97}
and from \citet{fontaine01} for helium-core or carbon/oxygen-core white dwarf
respectively. Figure~\ref{fig:simu} clearly shows that there is a lack of
systems containing white dwarfs with temperatures below $\sim$15,000K with
F-type companions (which would be found in the top-left of the plot), because
the F stars have also evolved and these systems are now double white dwarf
binaries. Figure~\ref{fig:simu} does suggest that we are missing a small number
of F stars with 15,000-25,000K white dwarf companions. Note that it is
possible for main-sequence stars with a large range of masses to all have
temperatures close to 6,200K (stars between 1.0 and 1.5{\MSUN} can have
temperatures close to 6,200K depending upon their surface gravity) and hence
there is a peak in the distribution shown in Figure~\ref{fig:simu} for stars
of this temperature.

\begin{figure}
\begin{center}
 \includegraphics[width=\columnwidth]{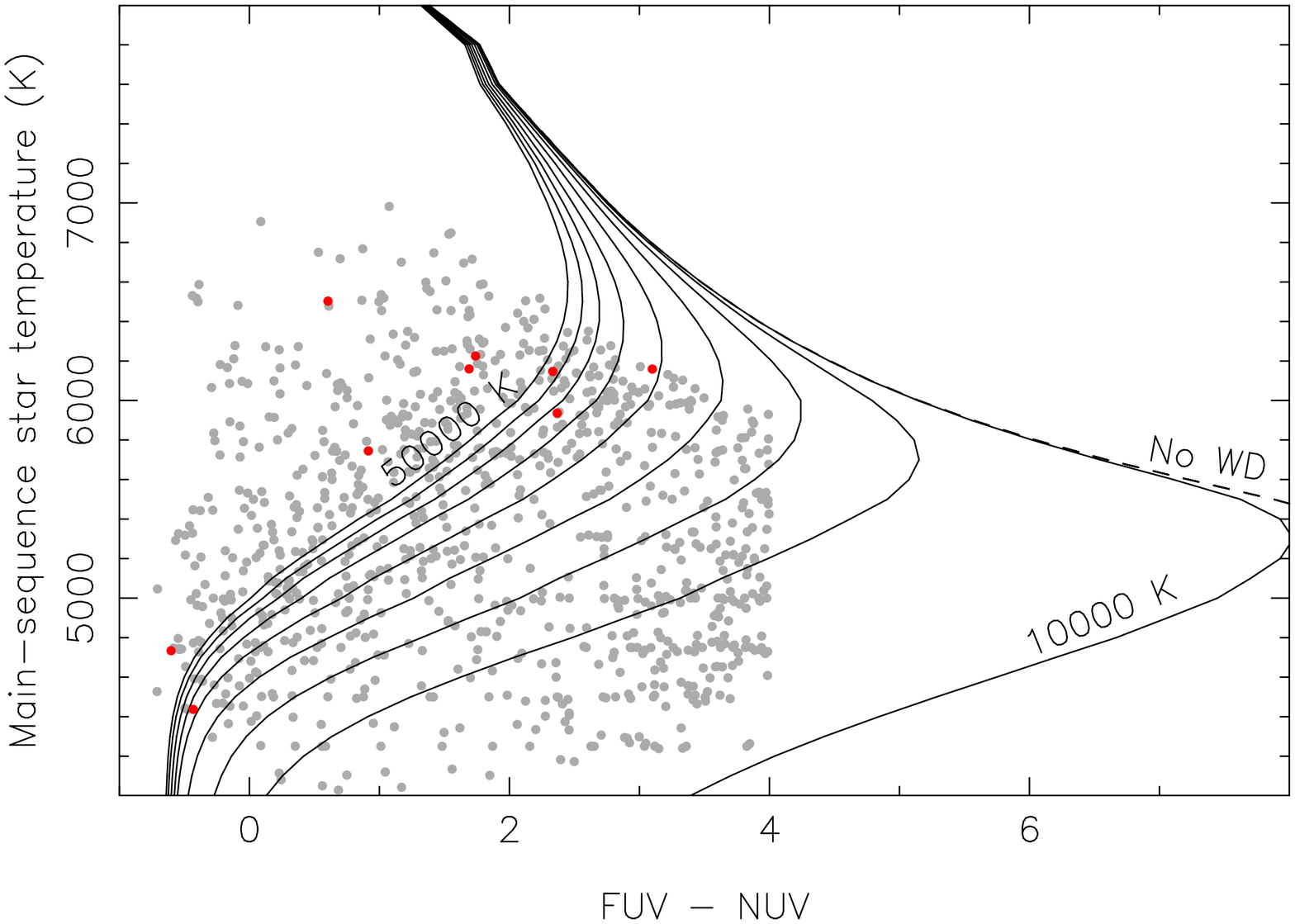}
 \caption{Same as Figure~\ref{fig:wdtemp} but with a $\log{g}=3.5$ metal poor
   ($\log{Z}=-3.0$) main-sequence star.}  
 \label{fig:wdtemp2}
 \end{center}
\end{figure}

\begin{figure}
\begin{center}
 \includegraphics[width=\columnwidth]{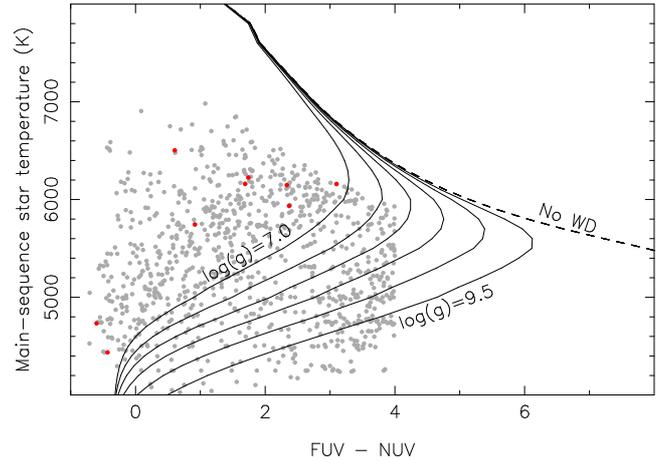}
 \caption{Same as Figure~\ref{fig:wdlogg} but with a $\log{g}=3.5$ metal poor
   ($\log{Z}=-3.0$) main-sequence star.}  
 \label{fig:wdlogg2}
 \end{center}
\end{figure}

Finally, the parameters of the main-sequence star (beyond its surface
temperature) also have an effect on the detectability of a white dwarf
companion. In Figure~\ref{fig:wdtemp2} and Figure~\ref{fig:wdlogg2} we show
the same plots as before, but with low surface gravity, metal poor
main-sequence stars, which have the bluest UV colours. In this case the
detectability is reduced in all circumstances and is particularly challenging
in the case of F stars, where even the hottest white dwarfs only produce an
excess of one magnitude, hence our sample is likely to contain only very few
such stars. Only 5 per cent of our UV-excess stars are metal poor
($\log{Z}<-1$) and all of these are K stars. Likewise, only 25 per cent of our
sample are low surface gravity stars ($\log{g}<4$), of which the majority
(65 per cent) are K stars and most of the rest (30 per cent) are G stars.

\section{Hubble Space Telescope UV observations} \label{sec:hst}

\subsection{Observations and their reduction}

We spectroscopically observed 9 UV-excess objects with \textit{HST} in order
to confirm that the excess is due to a white dwarf companion. Depending upon
the brightness of the target we either used the Space Telescope Imaging
Spectrograph (STIS) or the Cosmic Origins Spectrograph (COS). All observations
we obtained within program GO 13704 between October 2014 and July 2015 and
are summarised in Table~\ref{tab:hstobs}. For the STIS observations we used
the G140L grating centered on 1425{\AA} and for the COS observations we used
the G130M grating centered on 1291{\AA}. Each target was observed for one
spacecraft orbit. The data were processed using {\sc calstis} V3.4 and
{\sc calcos} V3.1. 

\begin{table}
 \centering
  \caption{Log of \textit{HST} observations.}
  \label{tab:hstobs}
  \begin{tabular}{@{}llll@{}}
  \hline
  Target             & Instrument & JD start & Exp (s) \\
  \hline
  CD-68\,2365        & STIS       & 2457182.2344 & 2712 \\
  LAMOST\,J1122+5236 & COS        & 2457015.3989 & 2309 \\
  LAMOST\,J1445+1236 & STIS       & 2457131.3753 & 2362 \\
  LAMOST\,J1528+4305 & COS        & 2456961.7038 & 2109 \\
  TYC\,6760-497-1    & STIS       & 2457031.7870 & 2381 \\
  TYC\,6917-654-1    & COS        & 2456987.8214 & 1977 \\
  TYC\,6996-449-1    & STIS       & 2457001.6869 & 2418 \\
  TYC\,7218-934-1    & COS        & 2457151.5144 & 2175 \\
  TYC\,7218-934-1    & COS        & 2457150.4747 & 2175 \\
  TYC\,9151-303-1    & COS        & 2457233.8558 & 2440 \\
  \hline
\end{tabular}
\end{table}

\subsection{Results}

\begin{figure}
\begin{center}
 \includegraphics[width=\columnwidth]{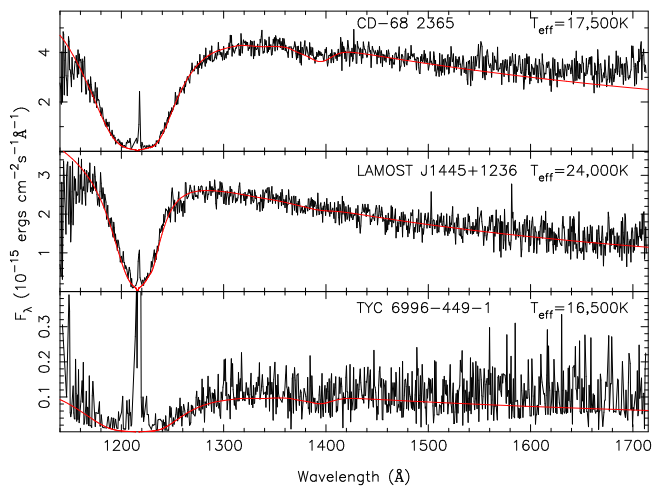}
 \caption{\textit{HST} STIS UV spectra of three of our UV-excess objects, clearly
   showing that the excess is the result of a nearby white dwarf. Also shown
   are the best fit white dwarf model spectra assuming a mass of
   0.6\MSUN.}  
 \label{fig:stis}
 \end{center}
\end{figure}

\begin{figure}
\begin{center}
 \includegraphics[width=\columnwidth]{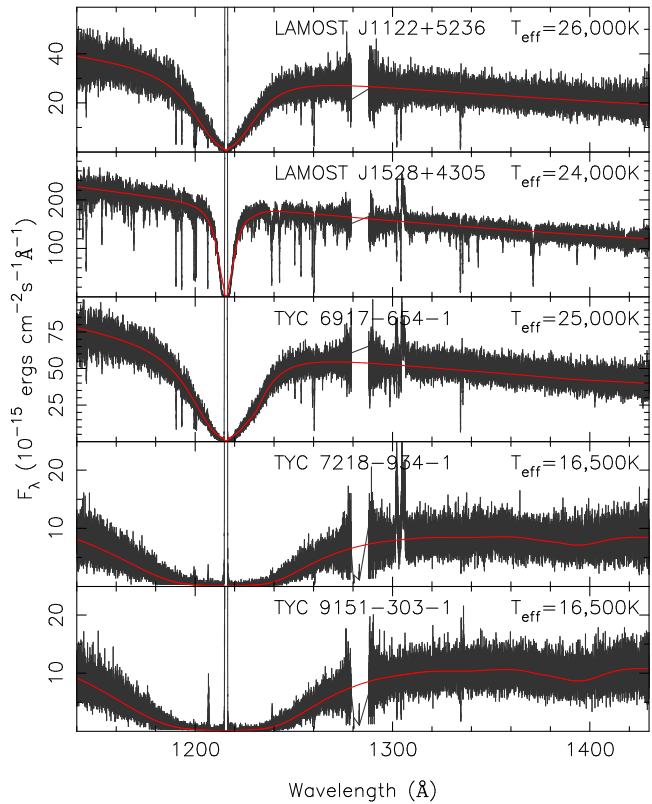}
 \caption{\textit{HST} COS UV spectra of five of our UV-excess objects, clearly
   showing that the excess is the result of a nearby white dwarf. Also shown
   are the best fit white dwarf model spectra assuming a mass of
   0.6\MSUN, with the exception of LAMOST\,J1528+4305, that cannot be fitted
   with a standard white dwarf model (see text). A $log{g}=5$ model is shown
   instead.}  
 \label{fig:cos}
 \end{center}
\end{figure}

We show the STIS and COS spectra in Figure~\ref{fig:stis} and
Figure~\ref{fig:cos} respectively. We confirm the presence of a white dwarf in
8 of the 9 observed systems. The analysis of TYC\,6760-497-1, which also
contains a white dwarf, was previously presented in \citet{parsons15} and is
not shown again. One system, LAMOST\,J1528+4305, appears to contain a
pre-white dwarf object. In this case the narrow Ly$\alpha$ line implies that
this object has a low surface gravity ($\log{g}\sim5$), which is too low for a
typical white dwarf. However, the blue continuum slope shows that the UV flux
originates from a hot star and is clearly not from the 4,700\,K main-sequence
star seen in the optical. Therefore, this hot object is fainter than the
main-sequence star in the optical and hence must be smaller. Therefore, we
suspect that this object could be either a hot subdwarf star or a pre-helium
white dwarf (similar to the objects seen in EL\,CVn type binaries
\citealt{maxted14}). The spectra of LAMOST\,J1122+523, LAMOST\,J1528+4305 and
TYC\,6917-654-1 also show narrow metal lines that may either be interstellar
or due to the white dwarf accreting wind material from the main-sequence star,
which may be an indication that some of these systems are close binaries
\citep[e.g.][]{parsons15}. All the spectra also show some geocoronal emission
in the core of the Ly$\alpha$ line and most of the COS spectra show geocoronal
airglow of O{\sc i} at 1310\AA. In at least one case, CD-68\,2365 in
Figure~\ref{fig:stis}, the main-sequence star appears to contribute a
noticeable flux towards the red end of the spectrum. 

We fitted TLUSTY/SYNSPEC \citep{hubeny95} white dwarf model spectra to each
of our \textit{HST} observations. With the UV data alone it is not possible to
simultaneously solve for the surface gravity, temperature and distance of the
white dwarf. Therefore, we fix the mass of the white dwarf at 0.6\MSUN (close
to the mean observed value for DA white dwarfs, \citealt{liebert05, falcon10})
and fitted the spectrum to get an estimate of the temperature and distance for
the white dwarf. The best fit models are overplotted in red in
Figure~\ref{fig:stis} and Figure~\ref{fig:cos} and detailed in
Table~\ref{tab:dists}. Note that in the case of LAMOST\,J1528+4305, where the
nature and hence radius of the compact star are unknown, we were unable to
determine a reliable distance estimate.We also estimated the distance to the
corresponding main-sequence stars in these systems by fitting their spectral
energy distributions (SEDs) using the virtual observatory SED analyzer (VOSA;
\citealt{bayo08}). We used archival optical data from the TYCHO and NOMAD
catalogues and infrared data from the 2MASS and WISE databases. We fitted the
SED with BTSettl models \citep{allard12} and kept the physical parameters of
the stars (effective temperature, surface gravity and metallicity) fixed at
the values from the RAVE and LAMOST databases and scaled the models to best
fit the SED, then used the resulting scale factor to determine the distance to
the star. The results are listed in Table~\ref{tab:dists}. 

\begin{figure*}
  \begin{center}
    \includegraphics[width=0.95\textwidth]{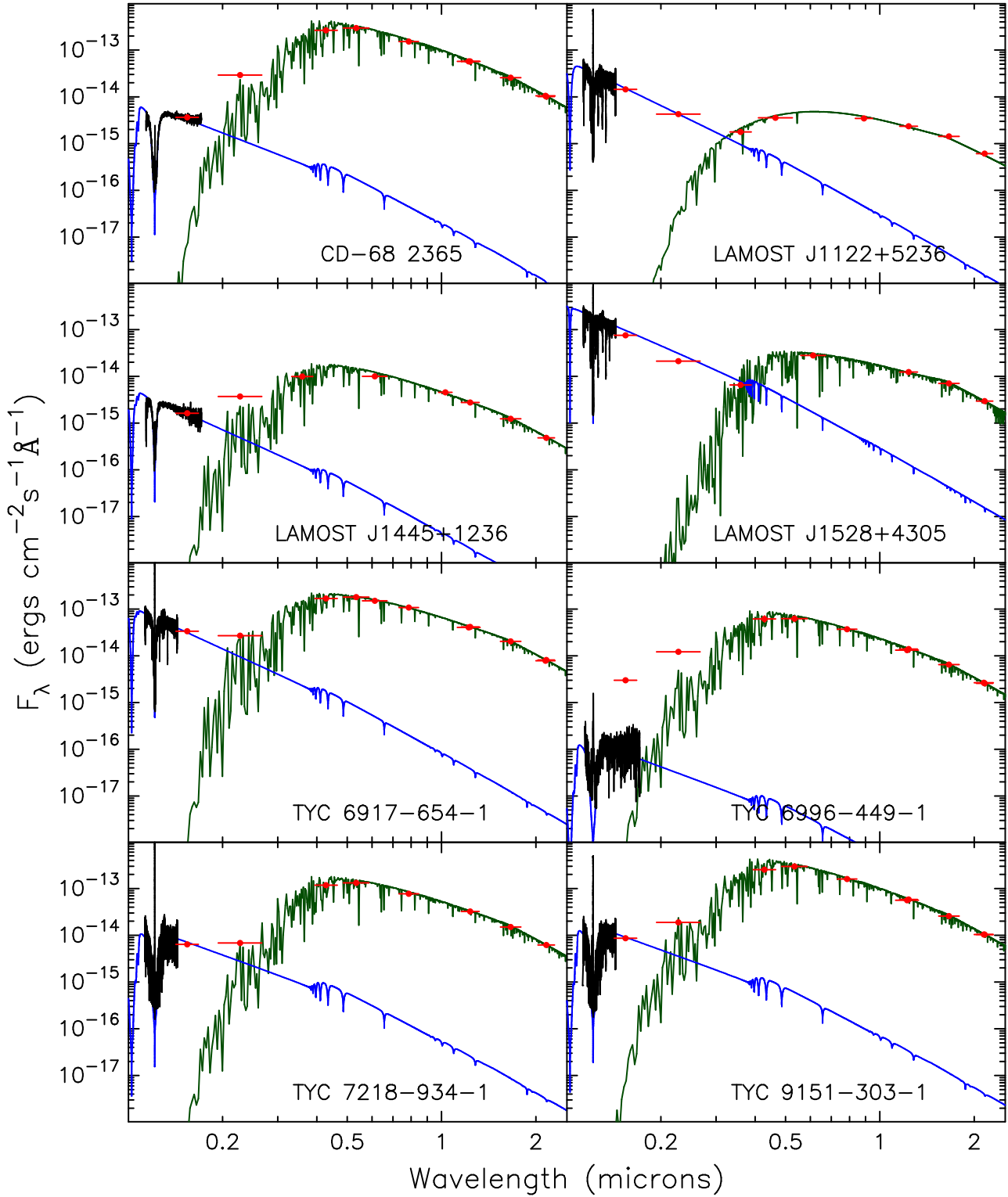}
    \caption{Spectral energy distributions for 8 UV-excess objects observed
    with \textit{HST}. Photometric measurements are shown in red (errors
    smaller than the symbol size) and are taken from GALEX (UV), TYCHO and
    NOMAD (optical) and 2MASS (near-infrared). The \textit{HST} spectra are
    shown in black. We also plot the best fit model spectra for the white
    dwarfs (blue lines) and main-sequence stars (green lines). For
    TYC\,6996-449-1 there is a substantial difference between the \textit{HST}
    flux and GALEX UV measurements. We believe that this is due to the white
    dwarf and main-sequence star being a spatially resolved binary (or the
    white dwarf is a background source) separated by a few tenths of an
    arcsecond. Since the \textit{HST} acquisition for STIS observations is
    done in the optical the slit may not have been properly centred on the
    white dwarf (but rather on the main-sequence star) resulting in slit 
    losses.}
  \label{fig:seds}
  \end{center}
\end{figure*}

\begin{table*}
 \centering
  \caption{Distance and physical parameter estimates to the white dwarfs 
    (assuming $\log{g}=8.0$) and main-sequence stars in the systems observed
    with \textit{HST}. Typical uncertainties in the white dwarf temperatures
    are $\pm500$K. The main-sequence star parameters are taken from the 
    RAVE/LAMOST pipelines. A reliable estimate of the distance to the hot 
    component in LAMOST\,J1528+4305 is not currently possible as the nature 
    of the star (and hence its radius) is unknown.}
  \label{tab:dists}
  \begin{tabular}{@{}lcccccc@{}}
  \hline
  Target & White dwarf & White dwarf & MS star & MS star & MS star & MS star \\
         & T$_\mathrm{eff}$ (K) & distance (pc) & T$_\mathrm{eff}$ (K) & $\log{g}$ & $\log{Z}$ & distance (pc) \\
  \hline
  CD-68\,2365        & $17,500$              & $271$          & $6159 \pm 66$  & $3.91 \pm 0.13$ & $ 0.00 \pm 0.10$ & $235 \pm 20$  \\
  LAMOST\,J1122+5236 & $26,000$              & $430$          & $4437 \pm 86$  & $3.95 \pm 0.34$ & $-0.71 \pm 0.39$ & $505 \pm 60$  \\
  LAMOST\,J1445+1236 & $24,000$              & $726$          & $6225 \pm 159$ & $4.20 \pm 0.46$ & $-0.61 \pm 0.40$ & $945 \pm 100$ \\
  LAMOST\,J1528+4305 & $24,000$              & -              & $4734 \pm 47$  & $2.99 \pm 0.45$ & $-0.09 \pm 0.20$ & $480 \pm 55$  \\
  TYC\,6917-654-1    & $25,000$              & $170$          & $6503 \pm 127$ & $4.01 \pm 0.21$ & $-0.25 \pm 0.13$ & $190 \pm 20$  \\
  TYC\,6996-449-1    & $16,500$              & $1400$         & $5937 \pm 95$  & $3.96 \pm 0.15$ & $-0.24 \pm 0.10$ & $425 \pm 40$  \\
  TYC\,7218-934-1    & $16,500$              & $165$          & $5746 \pm 95$  & $3.97 \pm 0.15$ & $-0.01 \pm 0.10$ & $190 \pm 15$  \\
  TYC\,9151-303-1    & $16,500$              & $151$          & $6161 \pm 95$  & $4.63 \pm 0.18$ & $ 0.13 \pm 0.11$ & $205 \pm 25$  \\
  \hline
\end{tabular}
\end{table*}

In Figure~\ref{fig:seds} we plot the UV to infrared SEDs of all eight systems
observed with \textit{HST} and the best fit white dwarf and main-sequence star
models. We do not show the RAVE spectra, since they cover a very small
wavelength range (near the Ca{\sc ii} infrared triplet), or the LAMOST
spectra, since the flux calibration of LAMOST spectra is relative
\citep{song12}. We found that in all but one case the distance derived for the
white dwarf and main-sequence star in reasonable agreement and therefore the
stars are likely in a physically associated binary, some of which may be
PCEBs, which we are aiming to identify in this sample (although additional
data are required to confirm this). However, we note that the distance
estimates to both stars are subject to systemic uncertainties. For determining
the distances to the white dwarfs, we fixed their masses at 0.6\MSUN, however,
if their actual masses are higher (lower) then their radii will be smaller
(larger) and hence they will be closer (further) than our estimates. Likewise,
the uncertainties on the physical parameters of the main-sequence stars lead
to a relatively large spread of possible distances. More accurate and precise
distances to the main-sequence stars will soon be available once the results
of the \textit{Gaia} mission are published. Only in one system,
TYC\,6996-449-1, is there is a clear discrepancy between the distance
estimates of the two stars, where the implied distance to the white dwarf is
more than three times that of the main-sequence star's. Forcing this white
dwarf to the same distance as the main-sequence star would require it to have
a mass in excess of the Chandrasekhar mass. However, inspection of
Figure~\ref{fig:seds} reveals that the flux of the \textit{HST} spectrum for
this object falls substantially below its measured GALEX magnitudes. We
believe that this is due to slit losses during the \textit{HST} observations,
caused by the fact that acquisition for STIS observations are done in the
optical, meaning that the slit is centered on the main-sequence star. If
there is a small offset between the main-sequence star and white dwarf
(i.e. if this system is a spatially resolved  binary or chance alignment),
then the 0.2{\arcsec} slit will not be properly centred on the white dwarf
component. This renders the white dwarf distance estimate unreliable and also
implies that this system is not a PCEB. However, re-scaling the COS spectrum
to fit the GALEX measurements reduces the distance to the white dwarf and
brings it into agreement with that of the main-sequence star, therefore, this
is likely to be a spatially resolved binary. Despite the fact that the
optical and GALEX  positional information for this target are coincident to
within 0.25\arcsec, the resolution and precision of the GALEX astrometry means
that we cannot exclude a spatial resolution of a few tenths of an arcsecond
between the main-sequence component and the white dwarf in any of our
systems. This means that, with the current data, we cannot easily distinguish
between PCEBs and main-sequence stars with nearby foreground/background white
dwarfs or wide binaries. 

\section{Discussion}

\subsection{Sources of contamination} \label{sec:contam}

Our sample of UV excess FGK stars contains close PCEBs (at least one has been
identified already, \citealt{parsons15}), wide binaries (with spatial
separations $\lesssim$0.25\arcsec) and chance alignments between FGK stars and
foreground or background UV bright sources (this includes white dwarfs, but
also hot subdwarfs and quasars). 

Several other types of object that could potentially contaminate our sample. 
The hot chromospheres of active stars can result in them possessing quite blue
UV colours. Additionally, any active star that flared during GALEX
observations may make it into our sample. Previous investigations of FGK stars
in SDSS determined that the strong chromospheres of some stars was likely the
cause of most UV excesses, particularly at later spectral types
\citep{smith14}. Recall that SDSS FGK stars are generally too distant to
detect any white dwarf companions and therefore the sample is dominated by
other types of UV-bright sources. Indeed, several stars in our sample are
known to be active and are listed as such in the SIMBAD database. However, the
vast majority of these are at the cooler end of our sample and hence are K
stars. 

As well as isolated active main-sequence stars, stars in close binary systems
can be spun-up and become active. Therefore, its possible that our sample is
contaminated by some active binaries. These are FGK plus M dwarf
binaries in which the unseen M dwarf has been spun-up and a flare from this
stars, or its strong chromosphere, is detected by GALEX. Since it appears that
most, if not all, M dwarfs in close binaries are active \citep{rebassa13},
this may be a source of contamination across our entire sample. Indeed,
several objects in our sample are listed as binary stars in SIMBAD (either
W\,UMa or RS\,CVn type variables). However, the fact that none of the systems
we observed with \textit{HST} were these types of binaries implies that the
contamination is relatively low.

Finally, we note that RR\,Lyrae stars can have NUV$-$FUV colours of between
$\sim$3$-$5 \citep{kinman14}, which places them within our cut. Generally,
most RR\,Lyrae stars are too faint to make it into our sample. However, while
there were no known RR\,Lyrae stars in our RAVE sample, there were 3 in our
LAMOST sample (since the LAMOST sample goes $\sim$2 mag deeper than RAVE), all
of which are at the fainter end of the sample (V>12). 

Given that all our \textit{HST} observations confirmed the presence of hot,
compact companions we conclude that there are few contaminating sources.
However, we note that our \textit{HST} observations mainly cover the
  larger UV excess objects and therefore more detailed observations are
required to better quantify this.

\subsection{Testing compact binary evolution}

This large sample of white dwarf plus FGK star binaries has the potential to
address several important questions relating to the evolution of compact
binaries and SN\,Ia formation. An initial step is to use spectroscopic data to
identify the PCEBs within the sample and measure their orbital
periods, a task that we have begun. High-resolution imaging can also be used
to identify the wide binaries and chance alignments.

One of the most interesting results from the handful of known PCEBs with
early-type components is that they can have a much larger range of orbital
periods after the CE stage \citep{zorotovic14}. While PCEBs containing M
dwarfs generally have periods below $\sim$10 days \citep{nebot11}, several
PCEBs containing early type stars have much longer periods. For example, the
white dwarf plus G5 binary KOI-3278 has a period of 88.2 days
\citep{kruse14}, while the white dwarf plus A8 binary IK\,Peg has a period of
21.7 days \citep{vennes98}. Conversely, there are still systems with very short
periods such as the 12.5 hour white dwarf plus K2 binary V471\,Tau
\citep{nelson70} or the 12 hour white dwarf plus F8 binary TYC\,6760-497-1
\citep{parsons15}. This suggests an additional source of energy is
  required during the CE phase in order to expel the envelope quickly enough
to prevent the two stars getting too close \citep{zorotovic10,rebassa12b},
although it is not clear what the source of this energy is
\citep{ivanova15}. This is particularly relevant to the 
evolution of double white dwarf binaries, since the two stars must emerge from
the first CE far enough apart to survive the second CE phase. So far, models
of the CE phase have struggled to simultaneously replicate the observed
populations of very close white dwarf plus M dwarf binaries and the population
of double white dwarf binaries, requiring two different formalisms to create
each population \citep{nelemans00,toonen13}. Therefore, a large,
well-characterised set of PCEBs with FGK components would be an extremely
valuable population for formulating a more complete description of the CE
phase itself, since it contains systems that emerged from the CE phase with
both very short and long orbital periods. 

This is also a particularly important subject for determining the progenitors
of SN\,Ia. Constraining the orbital period distribution of white dwarfs with
FGK companions (which will require additional spectroscopic observations)
will allow us to determine the fraction of systems following
the single and double degenerate formation channels. Moreover, our project has
already uncovered one system that will undergo a phase of thermal timescale
mass-transfer in the future \citep{parsons15} and hence, in this case, the
white dwarf will grow in mass. Given that it may be possible for
sub-Chandrasekhar mass white dwarfs to explode as SN\,Ia via the
double-detonation model it may even be that our sample contains some genuine
SN\,Ia progenitors.

\section{Conclusions}

We have presented a large population of 934 bright main-sequence F, G and K
type stars with UV excess colours that are likely the result of nearby white
dwarf companions. \textit{HST} spectra of nine of these objects has confirmed
that these are white dwarf plus main-sequence star binaries (in one case the
companion is likely a hot subdwarf star or pre-helium white dwarf) and there
appears to be very little contamination, at least among the larger UV
  excess objects. This previously undetected group of objects has the
potential to address a number of key questions relating to binary star
evolution and improve our understanding of the common-envelope
stage. Moreover, it will also constrain the ratio of binaries that undergo a 
single common-envelope event to those that will undergo two, with clear
implications for the study of type Ia supernovae.

\section*{Acknowledgements}

We thank the referee for helpful comments and suggestions. SGP and MZ
acknowledge financial support from FONDECYT in the form of grant numbers
3140585 and 3130559. This research was partially funded by MINECO grant
AYA2014-59084-P, and by the AGAUR. The research leading to these results 
has received funding from the European Research Council under the European
Union's Seventh Framework Programme (FP/2007-2013) / ERC Grant Agreement
n. 320964 (WDTracer). MRS thanks for support from FONDECYT (1141269) and
Millennium Science Initiative, Chilean ministry of Economy: Nucleus
P10-022-F. JJR acknowledges support by National Key Basic Research Program of
China 2014CB845700. Funding for RAVE has been provided by institutions of the
RAVE participants and by their national funding agencies. Guoshoujing
Telescope (the Large Sky Area Multi-Object Fiber Spectroscopic Telescope,
LAMOST) is a National Major Scientific Project built by the Chinese Academy of
Sciences. Funding for the project has been provided by the National
Development and Reform Commission. LAMOST is operated and managed by the
National Astronomical Observatories, Chinese Academy of Sciences.

\bibliographystyle{mnras}
\bibliography{wdfgk}

\appendix

\section{RAVE UV-excess targets}
\begin{table*}
 \centering
  \caption{Main-sequence stars from the RAVE survey with UV-excesses. The full
  table is available online.}
  \label{tab:ravetargs}
  \begin{tabular}{@{}lcccccccr@{}}
    \hline
SIMBAD Name             & RA           & Dec & FUV & NUV & V & T$_\mathrm{eff}$ (K) & $\log{g}$ & $\log{Z}$ \\
    \hline
UCAC3 131-28            & 00:00:44.521 & -24:41:16.62 & 22.246 & 19.969 & 11.960 & 4482 & 4.53 & -0.06 \\
2MASS J00005502-5738534 & 00:00:55.031 & -57:38:53.23 & 22.426 & 19.399 & 12.199 & 4719 & 4.35 &  0.29 \\
TYC 6992-827-1          & 00:01:13.398 & -33:27:06.73 & 18.014 & 16.944 & 11.053 & 5091 & 3.62 & -0.16 \\
UCAC2 17856213          & 00:03:08.614 & -32:49:52.76 & 20.843 & 18.390 & 12.560 & 5106 & 3.60 & -0.21 \\
TYC 1-890-1             & 00:04:21.129 & +01:09:14.46 & 19.248 & 18.563 & 11.141 & 5021 & 3.71 & -0.09 \\
BD-13 6521              & 00:07:09.421 & -12:22:21.98 & 14.877 & 14.205 & 10.962 & 5743 & 4.77 & -1.07 \\
TYC 5263-340-1          & 00:07:34.891 & -11:46:27.41 & 21.791 & 18.300 & 11.238 & 4646 & 4.75 &  0.60 \\
CD-44 16                & 00:08:44.585 & -43:42:24.73 & 20.064 & 16.185 & 10.519 & 5884 & 4.26 &  0.08 \\
TYC 8467-109-1          & 00:09:48.371 & -56:45:01.94 & 20.820 & 17.120 & 11.377 & 5287 & 4.68 & -0.04 \\
CD-26 60                & 00:14:12.099 & -26:05:50.77 & 15.432 & 13.741 & 10.230 & 6456 & 4.06 & -0.54 \\
TYC 9134-1979-1         & 00:15:12.269 & -68:50:58.49 & 21.805 & 19.077 &  8.760 & 5250 & 5.00 & -4.00 \\
TYC 6419-603-1          & 00:15:14.881 & -28:47:54.68 & 22.110 & 18.788 & 11.723 & 4574 & 4.28 &  0.72 \\
CD-30 57                & 00:15:19.606 & -29:46:15.89 & 17.703 & 16.183 & 11.523 & 5642 & 4.97 & -0.27 \\
LTT 128                 & 00:16:24.466 & -46:43:11.41 & 20.114 & 18.999 & 11.084 & 4348 & 4.86 & -0.04 \\
TYC 4670-766-1          & 00:17:32.176 & -07:21:13.52 & 18.686 & 15.975 & 10.391 & 5308 & 4.56 & -0.16 \\
UCAC2 10774801          & 00:19:01.431 & -48:53:36.44 & 21.615 & 18.729 & 12.190 & 4294 & 3.67 &  0.99 \\
TYC 6996-449-1          & 00:19:07.381 & -37:12:35.20 & 17.961 & 15.593 & 11.964 & 6162 & 4.04 & -0.56 \\
2MASS J00212706-1101097 & 00:21:27.068 & -11:01:09.70 & 21.164 & 17.889 & 12.130 & 5246 & 4.34 & -0.50 \\
2MASS J00214219-4140026 & 00:21:42.197 & -41:40:02.60 & 22.396 & 20.485 & 13.460 & 4000 & 4.78 & -0.50 \\
2MASS J00224832-6047096 & 00:22:48.324 & -60:47:09.68 & 21.500 & 19.142 & 12.150 & 4673 & 4.92 &  0.06 \\
...                     & ...          & ...          & ...    & ...    & ...    & ...  & ...  & ...   \\
    \hline
  \end{tabular}
\end{table*}

\section{LAMOST UV-excess targets}
\begin{table*}
 \centering
  \caption{Main-sequence stars from the LAMOST survey with UV-excesses. The
    full table is available online.}
  \label{tab:lamosttargs}
  \begin{tabular}{@{}lcccccccr@{}}
    \hline
    LAMOST Name & RA & Dec & FUV & NUV & $g$ or $r$(*) & T$_\mathrm{eff}$ (K) & $\log{g}$ & $\log{Z}$ \\
    \hline
    J0006+0247  & 00:06:40.63 & +02:47:04.9 & 20.637 & 18.941 & 15.48 & 6433 & 4.26 & -0.27 \\
    J0009+0420  & 00:09:48.87 & +04:20:43.5 & 18.869 & 18.653 & 14.90 & 5801 & 4.01 & -0.50 \\
    J0010+3445  & 00:10:38.07 & +34:45:34.3 & 19.435 & 18.633 & 13.70*& 5881 & 4.26 & -0.30 \\
    J0011+3416  & 00:11:43.58 & +34:16:16.7 & 20.331 & 18.741 & 12.97*& 5734 & 4.46 &  0.05 \\
    J0012+3424  & 00:12:33.67 & +34:24:31.8 & 15.865 & 14.850 & 10.90*& 6535 & 4.15 & -0.15 \\
    J0013+3135  & 00:13:38.16 & +31:35:12.3 & 18.698 & 17.501 & 11.32*& 5749 & 4.25 &  0.55 \\
    J0016+3002  & 00:16:41.20 & +30:02:38.4 & 19.261 & 15.496 &  9.83*& 5181 & 2.70 & -0.23 \\
    J0017+0521  & 00:17:42.15 & +05:21:10.0 & 18.985 & 19.211 & 18.41 & 4744 & 2.54 & -1.20 \\
    J0021+3342  & 00:21:22.99 & +33:42:37.1 & 19.637 & 17.064 & 10.49*& 4755 & 3.33 & -0.37 \\
    J0022+3322  & 00:22:39.48 & +33:22:16.7 & 19.754 & 19.199 & 13.25*& 5188 & 4.52 &  0.32 \\
    J0034+3951  & 00:34:16.74 & +39:51:05.8 & 20.066 & 19.513 & 15.30 & 4361 & 4.61 & -0.30 \\
    J0040+4057  & 00:40:07.81 & +40:57:56.7 & 22.396 & 19.676 & 15.62 & 6002 & 3.98 & -0.43 \\
    J0044+4138  & 00:44:33.80 & +41:38:28.5 & 21.751 & 20.241 & 16.75 & 5887 & 4.04 & -0.56 \\
    J0047+0319  & 00:47:05.92 & +03:19:54.8 & 17.338 & 16.414 & 15.96 & 5114 & 3.38 & -0.96 \\
    J0047+3944  & 00:47:38.86 & +39:44:11.6 & 19.480 & 17.767 & 14.87 & 6715 & 4.53 & -0.45 \\
    J0048+2706  & 00:48:04.18 & +27:06:57.6 & 18.385 & 16.414 & 12.48 & 6370 & 4.15 & -0.06 \\
    J0049+3919  & 00:49:27.53 & +39:19:18.9 & 20.679 & 19.128 & 16.04 & 6848 & 4.10 & -0.49 \\
    J0051+3925  & 00:51:03.61 & +39:25:35.0 & 22.054 & 21.498 & 15.73 & 4770 & 2.65 & -0.68 \\
    J0057+3625  & 00:57:46.23 & +36:25:07.0 & 20.039 & 18.819 & 15.34 & 6010 & 4.25 & -0.26 \\
    J0058+0235  & 00:58:23.64 & +02:35:56.7 & 18.157 & 14.235 & 11.64 & 5535 & 4.06 & -0.45 \\
    ...         & ...         & ...         & ...    & ...    & ...    & ...  & ...  & ...   \\
    \hline
  \end{tabular}
\end{table*}

\bsp
\label{lastpage}
\end{document}